\documentclass[a4paper,twocolumn]{esapub} 

\usepackage{graphicx}
\usepackage{times}
\usepackage{natbib}
\usepackage{amsmath}

\def\degr{\hbox{$^\circ$}}
\def\arcmin{\hbox{$^\prime$}}

\begin{document}

\pagestyle{empty}
\title{\large \bf Initial results from SECIS observations of the 2001 eclipse}

\author{\bf A.C.\ Katsiyannis$^{(1,2)}$, D.R.\ Williams$^{(3,2)}$, F.\ 
     Murtagh$^{(4,5)}$, R.T.J.\ M$^{\mathrm{c}}\!$Ateer$^{(6,2)}$,
     F.P.\ Keenan$^{(2)}$}

\affil{\it $^{(1)}$Department of Solar Physics, Royal Observatory of 
           Belgium, Avenue Circulaire -3- Ringlaan,B-1180, Belgium.,
           t.katsiyannis@oma.be}

\affil{\it $^{(2)}$Department of Pure and Applied Physics, 
           Queen's University Belfast, Belfast, BT7 1NN, U.K.,
           f.keenan@qub.ac.uk}

\affil{\it $^{(3)}$Mullard Space Science Laboratory, University College
           London, Holmbury St. Mary, Dorking, Surrey, RH5 6NT, U.K.,
           drw@mssl.ucl.ac.uk}

\affil{\it $^{(4)}$Department of Computer Science, Royal Holloway, 
           University of London, Egham, Surrey TW20 0EX, England, 
           fmurtagh@acm.org}

\affil{\it $^{(5)}$Observatoire Astronomique de Strasbourg, 11 rue de 
           l'Universit\'e, 67000 Strasbourg, France.}

\affil{\it $^{(6)}$ National Research Council, NASA Goddard Space 
           Flight Center, Greenbelt, MD 20771, U.S.A.,
           j.mcateer@grasshopper.gsfc.nasa.gov}

\maketitle

\thispagestyle{empty}

\abstract{SECIS observations of the June 2001 total solar eclipse 
were taken using an Fe~{\sc xiv} 5303 {\AA} filter. Automated tools
based on wavelet analysis was used to detect intensity oscillations on
various areas of the images. Statistical analysis of the detections
found in the areas covered by the moon and the upper corona allowed us
to estimate the atmospheric and instrumental effects on the detection
of intensity oscillations. An area of the lower corona, close to
Active Region 9513, was found with a statistically significant amount
of intensity oscillations with periodicity of $\sim$7.5$s$. The shape
of the wavelet transformation of those detections matches theoretical
predictions of sausage-mode perturbations and for the first time in the
SECIS project, second order oscillations were also detected.}

\section{\bf Introduction}

The magnetohydrodynamic (MHD) waves in corona loops have been
investigated as a possible cause of coronal heating by numerous
authors. Hollweg [1] first proposed a mechanism that involved MHD
waves dumping their energy into the solar corona through ion viscosity
and electrical resistivity, while other authors favoured magnetic
reconnection followed by current dissipation (for example Parker [2]).
Priest \& Schrijver [3] have recently published a review article
that contains all the theoretical attempts and their limitations.

MHD oscillations are usually divided in two main categories depending
on the presence or not of density perturbations. Magnetoacoustic MHD
waves are defined as the density, pressure and temperature
perturbations (which are in turn divided into slow and fast modes),
while the incompressible waves are called Alfv\'en and are also
divided into those with movements perpendicular to the magnetic field
and torsional. Roberts et al.\ [4] study a low-$\beta$ plasma, using
reasonable values for the solar corona approximations and predicted
that fast magnetoacoustic (sausage-mode) oscillations with frequencies
around 1 Hz can be excited in coronal loops. They also study the
signature of a wave train, created by a pulsation at $x=0$ observed at
a point $x=x'$. Much later, Nakariakov et al. [5] used numerical
analysis to confirm the initial study and provide a more accurate
simulation.

The Solar Eclipse Coronal Imaging System (SECIS) is an eclipse imager
that was developed specifically to detect corona oscillations, following
a long history of similar attempts (Koutchmy et al.\ [6], Pasachoff \&
Landman [7], Singh et al.\ [8], Cowsik et al.\ [9], Pasachoff et al.\
[10]). An Fe~{\sc xiv} 5303 {\AA} filter was used by the SECIS project
to observe the August 1999 and June 2001 total solar eclipses and the
results of the 1999 observations were reported in a sequence of papers
(Williams et al.\ [11], Williams et al.\ [12], Katsiyannis et al.\
[13]), while a detailed description of the instrument is provided by
Phillips et al.\ [14]. [11], [12] and [13] consistently detected
several periodicities in the range of 4-7 s, while [12] reported a
propagating wave train travelling through the corona loop with a
phase-speed of $\sim$2100~km~s$^{-1}$, making a fast-mode MHD wave as
the most likely explanation of the observed intensity perturbation.
By working on a different loop of the same active region [13] found
intensity oscillations just outside but fully aligned with visual
corona loops. Their detections were in the same frequency range and
have similar amplitudes as with [11] and [12]. After calculating the
physical characteristics of the loop (physical dimensions, density,
etc) and with reasonable assumptions regarding the strength of the
magnetic field, Roberts (private communication) confirmed that the
frequencies reported by the above authors ([11], [12] and [13]) are
well within the range predicted by [4].

Following the observations, reduction and data analysis of the 2001
eclipse were reported by Katsiyannis et al.\ [15]. In this paper we
will present further progress in the analysis of the data, a detailed
statistical analysis of the effects of the earth's atmosphere and some
preliminary detections of oscillations in the lower solar corona.

\section{\bf Observations}

An analytical description of the observations taken by SECIS during
the August 1999 total solar eclipse is presented by [14]. Although the
SECIS total solar eclipse observations of 2001 are described in detail
by [15], a briefer description of these observations is presented in
this section for clarity. In particular, the alterations made to the
instrument and some of the characteristics and specifications will be
presented below.

\begin{itemize}

\item The green Fe~{\sc xiv} 5303~{\AA} filter used in the previous 
eclipse was replaced with a broader, also Fe~{\sc xiv} 5303~{\AA}
filter, centred at practically the same line. The new filter has a
full-width-half-maximum (FWHM) of 5~{\AA} while the old filter was
just 2~{\AA}.

\item A metallic encloser was used to cover the optical elements of 
SECIS and reduce the amount of scattered light. The whole optical path
from the back of the Schmidt telescope to the charge-coupled device
(CCD) cameras was covered by this cover. A detailed diagram of the
instrument and the optical path covered can be found in [14].

\item Instabilities of the rotation axis speed of the heliostat have 
been observed by [14]. As a possible cause of those instabilities is
contamination of the driving mechanism of the heliostat, a cover was
produced that protected the area from dust. The analysis of the data
that followed showed no repeat of the 1999 instabilities on the 2001
observations.

\item The CCD cameras were cooled by newly installed fans, reducing the 
thermal effects.

\end{itemize}

The 21$^{\rm st}$ of June 2001 total solar eclipse was observed
following similar procedure to the August 1999 experiment. The
instrument was transfered in parts and was assembled on the the roof
of the Physics department of the University of Zambia, in Lusaka,
Zambia (Latitude: 15$\degr$ 20$\arcmin$ South; Longitude:28$\degr$
14$\arcmin$ East). The optical components of the instrument were
aligned using a pocket laser while the heliostat and the optical axis
of the instrument were aligned with the solar path using the standard
gnome technique. Distant objects were used for the focusing of the
various optic components.

During the eclipse weather conditions were good with practically no
wind or clouds. The cooling fans of the CCD cameras were switched on
early on the day of the eclipse but were switched off a few minutes
before totality to avoid vibrations to the instrument. The two
identical CCD detectors have 512 $\times$ 512 pixels, which combined
with the optics of the instrument provide us with an observable area
of $\sim$ 34$\times$34 arcmin$^{2}$ and a resolution of 4
arcsec~pixel$^{-1}$ (see [11] and references therein). Due to edge
effects and localised thermal effects on the CCD, an area of $\sim$
400 $\times$ 300 pixels was used for the data analysis. Also the
first 1000 frames of the data set were not used in any part of the
analysis described below as they were effected by light from the
photosphere during the start of the eclipse (an effect also known as
\lq\lq diamond rings effect'').  Having a previous knowledge of the
limitations of the CCD on the observable area and in line with
practice followed during the 1999 eclipse, the observations were not
centred at the centre of the disk but close to the North-East
limb. This location was chosen because of the appearance of NOAA
Active Region 9513 off the limb of the disk on the previous day. 8000
images were obtained during totality by the SECIS instrument with a
cadence of 39 frames per second, so the whole duration of the totality
was covered. The same area of the Sun was observed also by the Solar
and Heliospheric Observatory (SoHO) during totality so the physical
characteristics of AR 9513 can be determined.

Sky flat fields and dark frames (flats and darks for short) were
obtained the next morning. The same exposure time with the eclipse
observations was used for both the flats and the darks. For the flat
fields the heliostat was turned towards a featureless part of the sky,
while for the dark frames the apertures of the two lenses were closed
to f/22 and the cameras were covered completely with a black cloth.

\section{\bf Data Reduction and Analysis}

The data reduction software developed for the analysis of the 2001
eclipse data is described in detail by [15]. Since then various
improvements were made mostly to the way the images were aligned.
A brief description of the whole reduction and analysis is below 
and the differences with [15] will be explicitly mentioned.

\subsection{Image Alignment}

In a first stage dark current and sky flat subtraction was performed
in the standard way for astronomical images. 

The second phase involved alignment of the 8000 images to an accuracy
of one pixel.

The third step was the 8000 images to be aligned with an accuracy of 1
pixel. This was a different goal to what [15] tried to achieve as
later test revealed that the sub-pixel accuracy may introduce
artificial oscillations.  The first step of the alignment procedure
was to use the Sobel filter to calculate the edge of the moon disk and
then, assuming a constant lunar radius, a fit of the disk was produced
and compared with the edge fitted by the Sobel function. Those points
that were lying outside the fit by more than 3$\sigma$ were
rejected. In the next step we assumed that the centre of the moon was
moving in respect to the Sun with a constant velocity and the images
were corrected for this effect.  In the final stage further accuracy
was achieved by using an area of the 1000$^{\rm th}$ image containing
featureless parts of corona as well as a sharp transition from the
moon disk as a reference. This area was correlated with the equivalent
areas of the rest of the 8000 images and the pixel shifts for which
this correlation was maximum were considered the final shifts for the
alignment of the images. Unlike the procedure followed by [15], there
was no expansion of the correlation area.

\subsection{Wavelet analysis}

In line with previous work on SECIS ([11]-[13] and [15]) continuous
wavelet analysis was used to detect intensity oscillations. This
choice was made mainly because coronal loop oscillations are not
expected to necessarily last more than a few periods which meant that
a technique that produces the power of an oscillation over the whole
length of the experiment (like Fourier analysis) would be
unsuitable. As practically all the detections of intensity
oscillations done by SECIS have periods on the region of $\sim$5 to
8$s$, while the time series are either 40$s$ (1999 data set) or
205$s$, none of the detections reported by [11], [12] and [13] would
had been possible if Fourier analysis was used.

Torrence \& Compo [16] describe in detail the algorithm used and
provide a discussion of its benefits and application on various
scientific problems. A very brief description of the transformation
follows below for completeness.

\begin{equation*}
\psi(\eta)=\pi^{-1/4}\exp(i\omega_0\eta)\exp(\frac{-\eta^{2}}{2}),
\label{morlet}
\end{equation*}

\noindent where $\eta = t/s$ is the dimensionless time parameter, t
is the time, s the scale of the wavelet (i.e. its duration), $\omega_0
= s\omega$ is the dimensionless frequency parameter, and $\pi^{-1/4}$
is a normalisation term (see [16]).

\begin{figure}
\centering
\includegraphics[bb=200 0 504 520, width=3.5cm]
{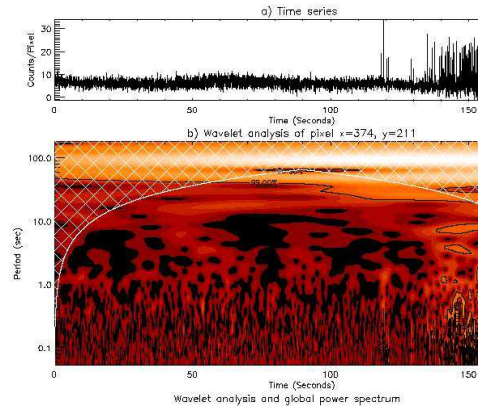}
\caption{Wavelet transform analysis of point x=374, y=211 of the aligned 
data set. Panel (a) contains the time series while (b) is the wavelet
transform. The contours in panel (b) highlight the areas where the
detected power is at the 99\% confidence level and the hatched area is
the cone-of-influence (COI).}
\label{wavelet_large}
\end{figure}

The results of the wavelet analysis described above applied to pixel
x=374, y=211 of the aligned 2001 data set are shown on Figure
\ref{wavelet_large}. Panel (a) is a plot of the time series of pixel 
x=374, y=211 of the aligned data. Panel (b) displays the power density
wavelet transform of the above time series with the lighter areas
representing the higher values. The hatched region marks the
cone-of-influence (COI) and represents the areas that suffer from edge
effects. Everything inside the COI should be discarded. More details
on the edge effects introduced by a finite time-series and how the
hatched area is affected by this can be found in [16] and the
references their in.  The contours of panel (a) surround the area
where the detected power exceeds the 99\% confidence level.

\section{\bf Automated Detection of Oscillations}

Software for the automated detection of intensity oscillations on the
SECIS data set was developed by [15] and was extensively used for this
study also. As with [15] this was done due to the large number of
pixels included in the data set. In total 5400 pixels were passed
through the scanning program for the determination of atmospheric and
instrumental effect and the detection of lower corona loop oscillation
(for more details see sections 5 and 6).

The following criteria were applied for the detections to be
considered valid. Both the choice of the mother wavelet and the
selection criteria were chosen for consistency with previous work by
[12], [13] and [15]. See [13] for a detailed discussion on the
establishment of these criteria.

\begin{itemize}

\item All values of coefficients falling within the COI are discarded.

\item Only areas with 99\% confidence level or higher are taken into 
account.

\item All oscillations lasting less than the time length of three 
periods were discarded.

\end{itemize}

A brief description of the software, written in IDL, follows:

\begin{enumerate}

\item For a given pixel of the data set the continuous wavelet 
coefficients of the time series was produced.

\item For the lowest periodicity the number of the first sample in 
time that is unaffected by the COI is determined. This sample is
called ${\rm t}$.

\item For the same periodicity the last sample that is unaffected by
the COI, ${\rm t'}$, was calculated.

\item The number of the sample that predates ${\rm t'}$ by three 
periodicities is determined. This sample is called ${\rm t}_{\mathrm
max}$

\item The confidence level of the sample ${\rm t}$ is calculated

\item If the confidence level of sample ${\rm t}$ is 99\% or higher, 
then the confidence level of the sample that is three periods later,
say ${\rm t_{+3}}$, is also extracted. Otherwise we go to step 8.

\item If the confidence level of ${\rm t_{+3}}$ is also 99\% or higher
the co-ordinates of the pixel, the current periodicity and ${\rm t}$
are recorded. In this case the algorithm goes to step 9. In this
point, it is assumed that if ${\rm t}$ and ${\rm t_{+3}}$ have both
confidence level of 99\% or higher, all samples between them will have
confidence level of 99\% or higher in the same periodicity. This
assumption is rarely wrong. All automated detections we have ever
inspected manually fulfil this hypothesis.

\item Go to the next ${\rm t}$ and repeat steps 5-7 until ${\rm t}$
becomes ${\rm t_{max}}$.

\item Go to next periodicity and repeat steps 2-8 until the 
periodicity reaches the limit of 70.9 $s$.

\item Move to the next pixel of the array and start again the entire 
procedure.

\end{enumerate}

70.9 $s$ is the upper limit of periodicities that can be detected in
the SECIS 2001 data set. This is because any oscillations with longer
periods do not have a long enough part of the time series outside
the COI to satisfy the criteria mentioned above.

\section{\bf Atmospheric and Instrumental effects}

\begin{figure}
\centering
\includegraphics[bb=0 140 504 590, width=7cm]{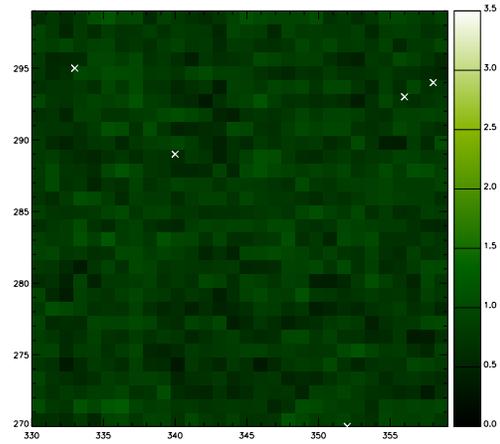}
\caption{A 30$\times$30 pixels area covered by the Moon. Out of the 
900 pixels, 5 were found to oscillate with a periodicity of 7-8
$s$. The pixels where those oscillations were found, are marked with
\lq x'. On average, the chance of a pixel of this area to be found to
oscillate is 0.56\%.}
\label{moon}
\end{figure}

Each frame of the SECIS eclipse observations can be divided into three
parts. The first is the area covered by the moon disk and contains
only signal from scattered line from the Earth's atmosphere. The
second part has signal mainly from the solar corona and the third part
is dominated by light coming from the outer corona.

\begin{figure}
\centering
\includegraphics[bb=0 130 504 590, width=7cm]{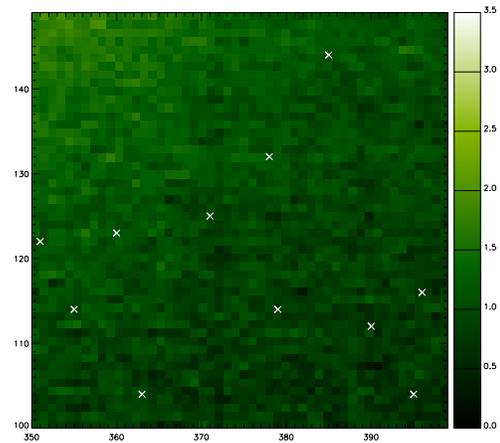}
\caption{A 50$\times$50 pixels area of the upper corona. Out of the
2500 pixels, 11 were found to oscillate with a periodicity of 7-8
$s$. On average, the chance of a pixel of this area to be found to
oscillate is 0.44\%.}
\label{upper}
\end{figure}

The different properties of these three areas are used in order to
estimate the atmosphere and instrumental effects to the detection of
intensity oscillations. The automated method and the criteria for
detecting oscillations, as defined in the previous sections, were
applied to a sample area of the moon and the outer corona. Figures
\ref{moon} and \ref{upper} contain the averaged, over the whole time 
sequence, images of an area of the moon and outer corona
respectively. The two axes of both images are the pixel coordinates of
the aligned data set and the green scale are the average pixel
values. Both areas were chosen to be close to AR~9513 but
significantly distant to the edge of the moon (for the disk sample)
and the edges of the useful area (for the outer corona sample). 

When applied to the sample moon area, for the detection of
oscillations with 7-8$s$ periodicity, the automated software produced
5 detections of oscillations out of 900 pixels. Since no signal from
the lower corona is detected in this area directly, all counts
generated will be either CCD read-out noise, or scattered atmospheric
light. When different ranges of periodicity are scanned by the same
software in the same area different pixels in random locations are
found to oscillate. This led us to believe that the moon detections
behave randomly and normal statistical procedures can be used to
determine the atmospheric and instrumental effects. Further
verification of this assumption is provided by the 50$\times$50 area
of the outer corona displayed in Figure \ref{upper}. Using the same
software, 11 oscillations were found in the same periodicity range,
making the average possibility of a pixel of this coronal area to
found with intensity oscillations 0.44\%. This is very close to 0.56\%
for the moon area reported above and confirms our assumption that the
intensity oscillation in the outer corona are too weak for SECIS to
detect them. Therefore any oscillations found in the upper are mostly
due to atmospheric and instrumental effects. To provide further
confirmation to this assumption more sample areas will be used from
both the lunar disk and outer corona areas and the possibility of a
pixel to oscillate will be calculated for each sample area
separately. The standard deviation of this measurements will provide
us with an estimation of how consistent the atmospheric and
instrumental effects are throughout the data set.

\section{\bf Detections of Oscillations around AR~9513}

\begin{figure}
\centering
\includegraphics[bb=0 200 504 550, width=7cm]{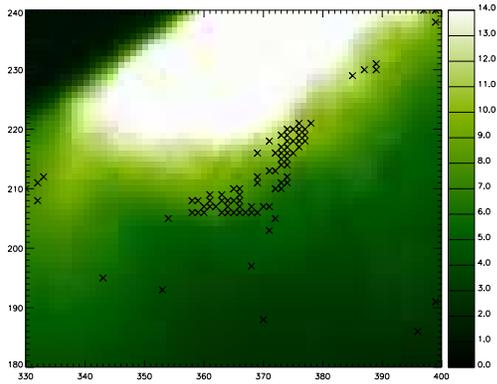}
\caption{The area outside AR 9513 where the intensity oscillations
with periodicity of 7-8$s$ were detected. All pixels with intensity
oscillations are marked with \lq x'.}
\label{lower_big}
\end{figure}

\begin{figure}
\centering
\includegraphics[bb=0 200 504 550, width=7cm]{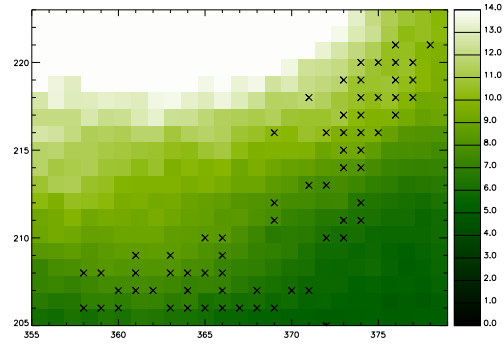}
\caption{The area of AR 9513 with the highest population of oscillations
with 7-8$s$ periodicity. This area is 25$\times$19 pixels in size
(475 pixels in total). Based on the statistics calculated by the
sample Moon and upper corona areas (see Figures \ref{moon} and
\ref{upper}), on average, there should be 2.4 detections due to
atmospheric or instrumental effects.}
\label{lower_small}
\end{figure}

Figure \ref{lower_big} presents an area of the lower corona close to
AR~9513 as observed by SECIS. Pixels with intensity oscillations that
passes the established criteria are marked with \lq x'. The detections
included in this Figure are, in their majority, due to plasma
intensity oscillation in the solar corona and not due to atmospheric
or instrumental effects for mainly two reasons:

\begin{enumerate}

\item The spatial distribution of the detected oscillations, unlike 
that of Figures \ref{moon} and \ref{upper}, is not random, but highly
concentrated in one area.

\item All 66 detections of interest are gathered in a very compact 
area.  Figure \ref{lower_small} contains all oscillations in a
25$\times$19 pixels frame of 475 pixels in total. Based on the
previous statistics an area of the moon or outer corona of the same
size should have 2.4 detections due to atmospheric or instrumental
effects.

\end{enumerate}

\begin{figure}
\centering
\includegraphics[bb=150 -20 504 500, width=3.5cm]
{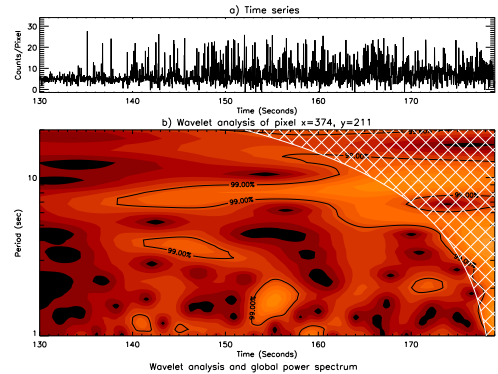}
\caption{Wavelet analysis of pixel x=374, y=211. This is a typical
sample of a detection of intensity oscillations in the area of Figure
\ref{lower_small}. The rest of the detections seen on Figure 
\ref{wavelet_large} and this image do not pass the criteria
established by [13].}
\label{wavelet_small}
\end{figure}

Figure \ref{wavelet_large} is a display the wavelet analysis of the
example point x=374, y=211 while Figure \ref{wavelet_small} is a
zoom-in of the area of Figure \ref{wavelet_large} that contains the
oscillations that pass the criteria established by [13]. 

Figure \ref{wavelet_small} has an interesting feature that appears in
some of the 66 detections reported here and has not been seen in the
1999 observations yet. Although not detected by the automated
software, as it does not satisfy the duration criteria established by
[13], there is an oscillation contemporary to the main detection at
exactly half the periodicity. This detection lasts for $\sim$12$s$ and
starts almost simultaneously with the main oscillation, 7.5 $s$. As
this secondary oscillation does not satisfy the duration criteria of
[13], we would normally ignore it, but there is a number of 
coincidences that make it worth reporting. Namely, the oscillation
starts practically simultaneously with the main one (and definitely
within the limits of the mother wavelet time resolution) and it has
exactly half of the periodicity of the main detection. Those two
properties make it a perfect candidate for identification as a second
order sausage-mode oscillation.

\section*{\bf Acknowledgements}

ACK acknowledges the Leverhume Trust for funding via grant
F00203/A. PPARC has funded part of this work. DRW and JMA thank DEL
and QUB for studentships. FPK is grateful to AWE Aldermaston for the
award of the William Penney fellowship.

\section*{\bf References}

\noindent 1.\ Hollweg, J.V., 1981, SolPhys, 70, 25 

\noindent 2.\ Parker, E.N., 1988, ApJ, 330, 474 

\noindent 3.\ Priest, E.R., Schrijver, C.J., 1999, SolPhys, 190, 1 

\noindent 4.\ Roberts B., Edwin P.M., Benz A.O., 1984, ApJ, 279, 857

\noindent 5.\ Nakariakov V.M., Arber T.D., Ault C.E., Katsiyannis A.C., 
Williams D.R., 2004, MNRAS, 349, 705

\noindent 6.\ Koutchmy, S., \v{Z}ug\v{z}da, Y.D. \& Loc\v{a}ns, V., 
1983, A\&A, 120, 185

\noindent 7.\ Pasachoff, J.M. \& Landman, D.A., 1984, Sol.\ Phys., 90, 
325

\noindent 8.\ Singh, J., Cowsik, R., Raveendran, A.V., Bagare, S.P., 
Saxena, A. K., Sundararaman, K., Krishan, V., Naidu, N., Samson, J.P.A., 
Gabriel, F., 1997, Sol.\ Phys., 170, 235

\noindent 9.\ Cowsik, R., Singh, J., Saxena, A.K., Srinivasan, R.
\& Raveendran, A.V., 1999, Sol.\ Phys., 188, 89

\noindent 10.\ Pasachoff, J.M., Badcock, B.A., Russell, K.D. \& Sea
ton, D.B., 2002, Sol.\ Phys., 207, 241

\noindent 11.\ Williams, D.R., Phillips, K.J.H., Rudawy P., Mathioudakis, 
M., Gallagher, P. T., O'Shea, E., Keenan, F. P., Read, P., Rompolt,
B., 2001, MNRAS, 326, 428

\noindent 12.\ Williams, D. R., Mathioudakis, M., Gallagher P.T., Phillips,
K. J. H., McAteer, R. T. J., Keenan, F. P., Rudawy, P., Katsiyannis,
A. C., 2002, MNRAS, 336, 747

\noindent 13.\ Katsiyannis, A. C., Williams, D. R., McAteer, R. T. J., 
Gallagher, P. T., Keenan, F. P., Murtagh, F., 2003, A\&A, 406, 709

\noindent 14.\ Phillips, K.J.H., Read, P., Gallagher P.T., Keenan, F. P.,
Rudawy, P., Rompolt, B., Berlicki, A., Buczylko, A., Diego, F.,
Barnsley, R.,Smartt, R.N., Pasachoff, J.M., Badcock, B.A., 2000, Sol.\
Phys., 193, 259

\noindent 15.\ Katsiyannis, A. C., McAteer, R. T. J., Williams, D. R., 
Gallagher, P. T., Keenan, F. P., 2004, SoHO, 13, 459

\noindent 16.\ Torrence \& C., Compo, G.P., 1998, Bull. Amer. Meteor. 
Soc., 79, 61

\end{document}